\documentclass{aa}  
\usepackage{graphicx}
\usepackage{amsmath}
\usepackage{natbib}
\usepackage{txfonts}
\newcommand{\der}[0]{\textrm{d}}

\begin{document}
   \title{Solving radiative transfer with line overlaps using Gauss Seidel algorithms.}


   \author{F. Daniel
          \inst{1}
          \and
          J. Cernicharo\inst{1,2}
          }

   \offprints{F. Daniel}

   \institute{Departamento de Astrofísica Molecular e Infrarroja, Instituo de la Estructure de la Materia,
              CSIC, Serrano 121, 28006 Madrid, Spain\\
              \email{[daniel,cerni]@damir.iem.csic.es}
           \and 
              Laboratoire d'Astrophysique de l'Observatoire de Grenoble, 
              414 rue de la Piscine, BP 52, 38041 Grenoble Cedex 9, France  
      }

   \date{Accepted 2008 June 13. Received 2008 April 30}

 
  \abstract
   {The improvement in observational facilities requires refining the 
    modelling of the geometrical structures of astrophysical objects. 
    Nevertheless, for complex problems such as line overlap in molecules showing
    hyperfine structure, a detailed analysis still requires a large amount of
    computing time
    and thus, misinterpretation cannot be dismissed due to an undersampling of  
    the whole space of parameters.}
   {We extend the discussion of the implementation of the Gauss--Seidel algorithm
    in spherical geometry and include the case of hyperfine line overlap.}
   {We first review the basics of the short characteristics method that is used to solve the radiative 
    transfer equations. Details are given on the determination of the Lambda operator in spherical geometry.
    The Gauss--Seidel algorithm is then described and, by analogy to the plan--parallel case, we see how to introduce it
    in spherical geometry. 
    Doing so requires some approximations in order to keep the
    algorithm competitive. Finally, line overlap effects are included.}
   {The convergence speed of the algorithm is compared to the usual Jacobi iterative schemes. The gain in the number 
    of iterations is typically factors of 2 and 4 for the two implementations made of the Gauss--Seidel algorithm.
    This is obtained despite the introduction of approximations in the algorithm.
    A comparison of results obtained with and without line overlaps for N$_2$H$^+$, HCN, and HNC
    shows that the $J=3-2$ line intensities are significantly underestimated in models where line
    overlap is neglected. }
   {}


   \maketitle
%

\section{Introduction}
Our understanding of the evolution of the molecular complexity in astrophysical
media relies both on the quality of the observations and on their interpretation.
The increasing spatial resolution that will be provided by the new generation of ground-- and space--based telescopes (Herschel, ALMA, JWST) will
enable finer geometrical structures to be infered in the objects under study. Hence,
the best way to interpret the origin of the molecular line emission and
to derive physical and chemical conditions is
to use 
non local radiative transfer calculations with realistic geometries adapted to the
angular resolution.
While the Sobolev approximation can be safely used when a large velocity gradient is present, interstellar clouds often show linewidths that are thermal in nature or that involve low--velocity fields, so the use of such an approximation is rather doubtful, and a widely used method in the field of molecular astrophysics
is the Monte--Carlo technique \citep{bernes1979}. This method is 
particularly attractive due to its conceptual simplicity and because it 
is easily implemented to treat multidimensional geometries. Nevertheless, an inherent drawback
of the method is that
it suffers from numerical noise and presents a slow convergence rate when optically thick lines
are modelled. Although line overlaps can be easily implemented \citep{gonzalez1993}, computing time becomes very
expensive when treating a case such as N$_2$H$^+$ where the two nitrogen atoms
contribute to the hyperfine structure.

Since the 90s, various techniques have been introduced to solve
the coupled set of radiative transfer and statistical equilibrium equations.
These methods lead to substantial gains in term of calculation speed 
and accuracy of the solutions. Among them, the preconditioning of statistical 
equilibrium equations entails drastic improvement in the convergence 
properties of the iterative scheme for problems with large optical depths \citep[see][]{rybicki1991,rybicki1992}. A crucial issue is that the Lambda iteration scheme does not always converge to the true solution for large optical depths \citep[see][]{mihalas1978,auer1991}.
Another progress consisted in the introduction by \citet{trujillo1995} of the Gauss--Seidel algorithm 
for the solution of radiative transfer problems.

The description of this method was first made for the case of one-dimensional plane-parallel
stellar atmospheres \citep{trujillo1995,paletou2007} and was further
generalised to the case of two and three-dimensional radiative transfer problems with
multilevel atoms in Cartesian coordinates \citep{bendicho1997,leger2007} and to the
case of scattering line polarization \citep{trujillo1999}.
It was then described \citep{asensio2006} and used \citep{cernicharo2006}
in spherical geometry. Although the geometry of interstellar clouds cannot always be assumed to be as simple as spherical, it enables a first guess more appropriate than plan--parallel geometry in order to model centrally condensed clouds.
In this article, we give more details on the implementation in spherical geometry,
and we extend the discussion to the case of line overlap. We find that
in spherical geometry, the requirements of the original Gauss--Seidel algorithm 
cannot be fulfilled without increasing the computational time (even in the
case of isolated lines).
To address this problem, we introduced several
approximations into the algorithm. Interestingly, despite 
these approximations, the usual increase in reaching convergence 
in terms of number of iterations persists. Moreover, 
these approximations do not entail false convergence, in comparison to results obtained
within the $\Lambda$ iteration scheme.

Several molecules used to trace the physical conditions in molecular clouds
harbour a hyperfine structure that complicates the analysis of the observed emission \citep[ see e.g.][]{daniel2007}.
Indeed, the influence of the overlap is usually not considered when interpreting observations 
despite the fact that for a molecule with hyperfine structure, the line blending increases with the rotational quantum number. For high rotational transitions, all the hyperfine lines are usually overlapped in astrophysical objects. 
Previous works have shown that, for HCN, line overlaps play an important role in the pumping of molecular levels and have to be considered in order to interpret
the relative intensities of hyperfine transitions \citep[cf.][]{guilloteau1981,gonzalez1993}. One goal of this paper is to extend the discussion to other molecules, i.e. HNC and N$_2$H$^+$, and to derive the general radiative transfer effects that are common to the different species.

This article is organised as follows. In section \ref{section:th1}, we first 
overview the short characteristics (SC) method. In section \ref{theorie:ALI}, 
the main equations used 
in the preconditioning of statistical equations are given. 
In section \ref{theorie:GS}, the implementation of the Gauss--Seidel algorithm
in spherical geometry is discussed.
In section \ref{app_num}, a test case
for H$_2$CO gives the main properties of the various algorithms.
Finally, in section \ref{app_LO}, the influence of introducing line overlap in
the calculations is discussed for the N$_2$H$^+$, HCN, and HNC molecules.


\section{Theory} \label{section:Theory}

\subsection{Short characteristics in spherical geometry} \label{section:th1}

\begin{figure}[h!]
   \centering
   \includegraphics[scale=0.3]{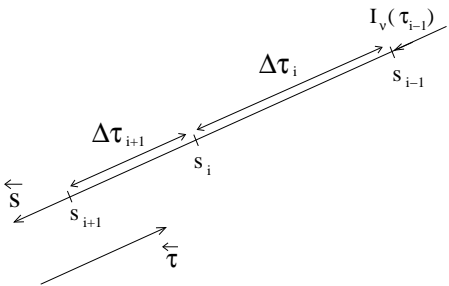}
      \caption{Transfer along a short characteristic.}
         \label{fig1}
\end{figure}
To solve the transfer equation, we use the SC method
introduced by \citet{olson1987}. Here, we briefly overview the method and give 
the main equations in order to introduce quantities that will be used in the following discussion.

If one considers the radiation propagation along 
a ray that is discretized (cf. Figure \ref{fig1}), 
the specific intensity between two consecutive points varies as \citep{olson1987}
\begin{eqnarray}
I_{\nu} (\tau_i) & = & I_{\nu} (\tau_{i-1}) \, \textrm{e}^{-\Delta \tau_i} + \int_{\tau_{i-1}}^{\tau_{i}} S_{\nu}(\tau) \, \textrm{e}^{\tau_i-\tau} \, \der \tau \nonumber \\
\Leftrightarrow  \quad I_i & = & I_{i-1} \,  \textrm{e}^{-\Delta \tau_i} + \Delta I_{i}
\qquad .
\label{eq1}
\end{eqnarray} 
Assuming a parabolic trend for the source functions with respect to the opacity of the line,  we can derive the intensity analytically as follows:
\begin{eqnarray}
\Delta I_i  = 
\alpha_i \, S_\nu (\tau_{i-1}) + \beta_i \, S_\nu (\tau_i) + \gamma_i \, S_\nu (\tau_{i+1})
\label{eq2}
\end{eqnarray}
where the expressions of the coefficients $\alpha_i$, $\beta_i$, and $\gamma_i$ are given by \citet{olson1987}
and depend on $\Delta \tau_i$ and $\Delta \tau_{i+1}$. These latter quantities are derived assuming that the absorption
coefficients $\kappa_\nu(s)$ vary linearly between two consecutive grid points.

\begin{figure}
   \centering
   \includegraphics[scale=0.4]{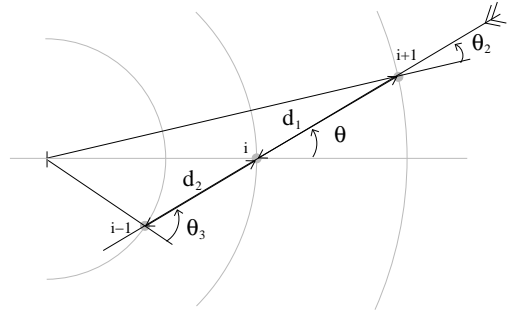}
      \caption{Geometric quantities in spherical geometry.}
         \label{fig2}
\end{figure}

In spherical geometry, the radiation field is fully characterised introducing two variables, $r$ and $\theta$.
Numerically, one has to adopt a discretized space for these two variables. At a given radius $r_i$ and for an angle 
$\theta_i$, the specific intensity is known within the SC scheme from the geometrical quantities represented in Fig. \ref{fig2} . In the case $\theta_i < \pi/2$, we find
\begin{eqnarray}
\begin{array}{lll}
d_{1,i} & = &\displaystyle - r_i \cos \theta + \sqrt{r_{i+1}^2-r_i^2 \sin^2 \theta_i} \vspace{0.2cm } \\
\theta_{2,i}  & = &\displaystyle \textrm{arcos} \left( \frac{d_{1,i}^2+r_{i+1}^2-r_{i}^2}{2 \, d_{1,i} \, r_{i+1}}\right) \vspace{0.2cm } \\
d_{2,i} & = & \displaystyle r_i \cos \theta_i - \sqrt{r_{k}^2-r_i^2 \sin^2 \theta_i} \vspace{0.2cm }  \\
\theta_{3,i} & = & \displaystyle \textrm{arcos} \left( \frac{r_i^2-r_{k}^2-d_{2,i}^2}{2 \, d_{2,i} \, r_{k}}\right) \vspace{0.2cm } \\
\end{array}
\\ \textrm{with} \qquad k \equiv \left\{
\begin{array}{lcl}
 i+1 &\quad \textrm{if} \quad &\theta_i = \pi/2 \vspace{0.3cm} \\
 i    &\quad \textrm{if} \quad &\theta_i \in \, ]\pi/2; \xi_{i,i-1}[ \vspace{0.3cm} \\
i-1 &\quad \textrm{if} \quad &\theta_i \geq \xi_{i,i-1} \vspace{0.3cm } \qquad \qquad .\\
\end{array} \right. \\
\end{eqnarray}
The angle $\xi_{m,n}$ is defined according to
\begin{eqnarray}
\xi_{m,n} = \pi - \textrm{arccos} \sqrt{1-  \left( \frac{r_{n}}{r_m} \right)^2}
\label{angle}
\end{eqnarray}
and stands for the angle sustained by the layer of radius $r_n$ seen from the layer of radius $r_{m}>r_n$.

The quantities $d_{1,i}$ and $d_{2,i}$ are used to derive $\Delta \tau_i$ and $\Delta \tau_{i+1}$. The angle $\theta_{2,i}$ is used
to interpolate the incoming intensity. Moreover, the angle $\theta_{3,i}$ is used in the determination of $\Delta \tau_{i+1}$ when 
a radial velocity field is introduced.
For the case $\theta_i > \pi/2$, the above formulae also apply, using the angle $\pi-\theta_i$ and then inverting respectively $\theta_{2,i}$ with
$\theta_{3,i}$ and $d_{1,i}$ with $d_{2,i}$.

\subsection{Accelerated Lambda Iteration}\label{theorie:ALI}
 
Following \citet{olson1986}, rather than the lambda iteration scheme (LI), we use an alternative scheme known as the accelerated (or approximated) lambda iteration (ALI) since this one presents better convergence properties. In that scheme, the specific intensity obeys
\begin{eqnarray}
I_{\nu \, \Omega} 
 & = & \left ( \Lambda_{\nu \, \Omega} - \Lambda_{\nu \, \Omega}^* \right)[S^+_\nu] + \Lambda_{\nu \, \Omega}^* [S_\nu] + \mathcal{J}^+_{\nu \, \Omega}
\label{eqALI}
\end{eqnarray}
where $\Lambda_{\nu \, \Omega}$ stands for the exact Lambda operator and $\Lambda_{\nu \, \Omega}^*$ for an approximated Lambda operator (ALO). In this expression, the indexes correspond to the frequency $\nu$ and to 
the direction of propagation $\Omega \equiv (r,\theta)$. 
The quantities indexed by $^+$ are determined using the level populations obtained at the previous iteration.  
The value $\mathcal{J}^+_{\nu \, \Omega}$ corresponds to the external radiation field which is transmitted to the point 
under consideration. 
From Eq. \ref{eqALI} the LI scheme would be obtained letting $\Lambda_{\nu \, \Omega}^* \equiv 0$.

Several studies aiming at determining the most effective ALO to be used so that the convergence is 
reached in the fastest way have been conducted. Non diagonal Lambda operators were found to be efficient in reducing the number 
of iterations but at a computational cost per iteration higher than for a diagonal Lambda operator. Finally, the use of a 
diagonal ALO seems to lead to the fastest convergence rates \citep[see e.g.][]{puls1988,olson1986}.

When we use the diagonal part of the full lambda operator for $\Lambda_{\nu \, \Omega}^*$, we find that
instead of solving the usual statistical equilibrium equations (SEE), one has to solve the system \citep{rybicki1991}
\begin{eqnarray}
0 & = & \sum_{j>i} [ n_j A_{ji} (1-\overline{\Lambda}^*_{ji}) - (n_i B_{ij} - n_j B_{ji}) \overline{J}_{ji}^{\, \textrm{eff}} ] \nonumber \\ & & 
   - \sum_{j<i} [n_i A_{ij}(1-\overline{\Lambda}^*_{ij}) - (n_j B_{ji} - n_i B_{ij}) \overline{J}_{ij}^{\, \textrm{eff}} ] \nonumber \\ &  & 
   + \sum_{j \neq i} [ n_j C_{ji} - n_i C_{ij} ]
\label{geom:eq_stat4}
\end{eqnarray}
with
\begin{eqnarray}
\overline{J}_{ij}^{\, \textrm{eff}} = \overline{J}_{ij}^+ - \overline{\Lambda}^*_{ij} \times S^+_{ij}
\qquad .
\end{eqnarray}
In the above equation, $S_{ij}$ is the usual frequency--independent source function for an isolated molecular line. 
Th term $ \overline{J}_{ij}^+$ corresponds to the radiation field averaged over frequency and angle. 
The above expressions apply with or without any source of background continuum radiation. When there
is a source of background continuum radiation, the source function is given by \citep{rybicki1991}
\begin{eqnarray}
S{_\nu} = r_{ij} S_{ij} + (1-r_{ij}) S_c
\end{eqnarray}
with
\begin{eqnarray}
r_{ij} = \frac{\kappa_{ij}(\nu)}{\kappa_{ij}(\nu)+\kappa_c(\nu)} \qquad ,
\end{eqnarray}
where $\kappa_{ij}(\nu)$ is the absorption coefficient due to the molecular transition and $\kappa_c(\nu)$ 
is the continuum absorption coefficient. 
The averaged Lambda operator is given by
\begin{eqnarray}
\overline{\Lambda}^*_{ij} = \frac{1}{4\pi} \int_{\Omega} \int_{\nu} \phi_{ij}(\nu) \, \Lambda_{\nu \, \Omega}^* \, r_{ij}^+ \, \der \nu \, \der \Omega  \qquad ,
\end{eqnarray}
where $\phi_{ij}(\nu)$ is the intrinsic line shape function.

\subsubsection{Overlap of hyperfine lines}\label{theorie:LO}

The overlap between hyperfine lines is treated  as 
described by \citet{rybicki1992}. Here we give a brief overview of 
the main equations.
In the case of overlapping lines, the source function without background continuum radiation 
is given by
\begin{eqnarray}
S_\nu = \frac{1}{\kappa(\nu)} \sum_{\substack{ ij \\ i>j}} \kappa_{ij}(\nu) \, S_{ij} \qquad ,
\end{eqnarray}
where $S_{ij}$ is the usual frequency independent source function encountered
in the case of an isolated line and $\kappa_\nu$ is the total
absorption coefficient at frequency $\nu$ given by 
\begin{eqnarray}
\kappa(\nu) =  \sum_{ \substack{ ij \\ i>j}} \kappa_{ij}(\nu) \qquad .
\end{eqnarray}
Within the SC scheme, these quantities are used to perform the analytical integration 
along a characteristic using Eqs. \ref{eq1} and \ref{eq2}.
As explained in \citet{rybicki1992}, the preconditioning of the statistical equilibrium 
equations can be made in different ways. In the present work, we use the preconditioning with the same transition. This corresponds to eq. 2.23 of \citet{rybicki1992}. Rewriting this latter equation in terms of the Lambda operator rather than
the $\Psi$ operator which acts on emissivities and considering that a diagonal ALO acts like a simple multiplication leads to the same 
form of the preconditioned statistical equilibrium equations as given by Eq. \ref{geom:eq_stat4}. The only difference is that 
the averaged diagonal ALO is now given by

\begin{eqnarray}
\overline{\Lambda}^*_{ij} = 
\frac{1}{4\pi} \int_{\Omega} \int_{\nu} \phi_{ij}(\nu) \, \frac{\kappa_{ij}(\nu)}{\kappa(\nu)} \, \Lambda_{\nu \, \Omega}^* \, \der \nu \, \der \Omega \qquad .
\end{eqnarray}

We note that the full preconditioning strategy might be more adapted to the case of the hyperfine structure where lines are close in frequency. Thus, the present choice is motivated by the SEE equations remaining unchanged and are still given by Eq. \ref{geom:eq_stat4}. 

In the appendix, the way the Lambda operator is constructed is presented and the differences with the plan--parallel case are emphasised.

\subsection{Gauss-Seidel}\label{theorie:GS}

The Gauss--Seidel (GS) iterative scheme was introduced in radiative transfer by \citet{trujillo1995}.
The superiority of this algorithm was then shown, by comparison to Jacobi--based ones,
in order to rapidly reach convergence even for the complex case of multilevel atoms in two and three-dimensional
Cartesian geometry \citep[see][]{bendicho1997}. Recently, a detailed description of the implementation of the GS
algorithm for multilevel atomic (or molecular) case was presented by \citet{paletou2007}.
So far, the GS algorithm has been extensively described in plan--parallel geometry.
The description of the implementation in spherical geometry has been done \citet{asensio2006},
and here we extend the discussion made in this article. Before,
we will review the main attributes required by the algorithm.

At each iteration, the grid is first swept from the outermost layer until 
the innermost one (say, from $k=N \to 1$ ), and during this process, part of the averaged radiation field is computed, i.e. for $\theta \in [0;\pi/2]$. Then, once the innermost layer is reached, the grid is swept in the other direction. This leads to the computation of the averaged radiation field that corresponds to $\theta \in [\pi/2;\pi]$. Within the Jacobi iterative scheme, all the grid is swept until reaching the outermost layer, and then statistical equilibrium equations are solved. Within the GS iterative scheme, once the averaged radiation field is known for a layer, the statistical equilibrium equations are 
solved for it.  At this stage, various updates have to be made to account for the newly derived populations.
Using parabolic interpolation, this implies correcting both the outward and inward radiation field respectively, for the 
current and the next grid point. These corrections are made so that when deriving
the populations for a layer $k$, the populations considered are those obtained during the current iteration 
for the layers $i<k$ and those corresponding to the previous iteration for the layers $i \geq k $, noted 
$\left\{ n_j^{new} \right\}_{i<k}$
and  $\left\{ n_j^{old} \right\}_{i \geq k}$. \\

We now describe how the GS algorithm should be implemented in spherical geometry
and how we have implemented it in practice. In what follows, we note $I_k^{in}$, the intensity 
$ I_k(\theta)$ for $\theta \in [0;\pi/2]$ and $I_k^{out}$ the intensity 
$I_k(\theta)$ for $\theta \in [\pi/2;\pi]$.

\subsubsection{Upward inversion} \label{upward_inv}

Let us consider first that the grid has been swept from the outermost layer to the innermost one. This leads to the computation of $\left\{ I_k^{in} \right\}_{k\in[1;N]}$. Once the innermost layer is
reached, we can compute the intensities $I_1^{out}$  since
the incoming intensities are given by $I_1^{in}$  
(If there is a central source of continuum radiation, the intensities are an input parameter.). We then know 
$\overline{J}$, and we can invert the statistical equilibrium equations for this layer. 
Before going to the point $k=2$, there are three corrections that in principle have to be done to conform with the requirements of the original GS algorithm:
\begin{itemize}
\item an update of $I_1^{in}$  which is computed with $n_1^{new}$ and $n_2^{old}$
\item an update of $I_1^{out}$ 
\item an update of $I_2^{in}$ computed with $n_1^{new}$ and $n_2^{old}$ and 
           $n_3^{old}$  \quad .
\end{itemize}
Note that the first correction should be performed because the intensities $I_1^{in}$ are used to derive the incoming intensities when determining $I_1^{out}$. However, this correction is not necessary in practice since it influences the convergence rate just slightly (cf. section \ref{app_num}).
Once these corrections are done, we go forward in the propagation. 

\begin{figure*}
   \centering
   \includegraphics[scale=0.6,angle=270]{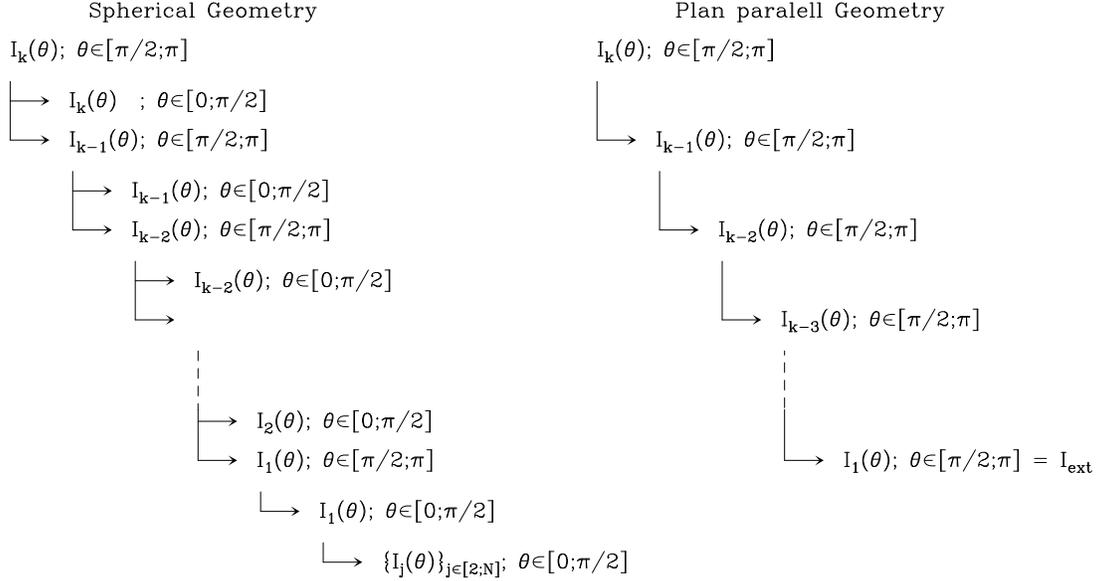}
      \caption{Dependence of the intensities.}
         \label{fig3}
\end{figure*}

Suppose we reach the point $k$ where new populations are obtained. Let us now consider the corrections that are needed to have an exact GS iterative scheme. Since new populations have been obtained for the layer $k$, a requirement of the algorithm is that, at the point $k+1$, the populations should be determined with $\{n^{new}_j \}$ for $j\in[1;k]$ and $\{n^{old}_j \}$ for $j\in[k+1;N]$. Now, consider the way $I_k^{out}$ depends on the various populations and so, through the dependance on the incoming intensities.
For this purpose, we refer to Fig. \ref{fig3}, which shows the dependencies of intensities. We see that to 
obtain the  intensities $I_k^{out}$, it is necessary to know the inward intensities $I_k^{in}$, as well as the outward intensities $I_{k-1}^{out}$, which in turn depend on $I_{k-2}^{out}$, etc. Finally, we see that fully accounting 
for the new population $n_k^{new}$ in
 the computation of $I_k^{out}$ would require first to perform the propagation along the inward direction, leading to the updated set
$\{I_j^{in}\}_{j\in[1;k+1]}$. Then, the propagation would be done again along the outward direction so that the set $\{I^{out}_j\}_{j\in[1;k]}$ would effectively be given considering the new populations $n^{new}_{k}$. 
In practical applications, such a process would lead to computational times longer than those obtained within the Jacobi iterative scheme.  Moreover, it should be noted that this problem does not arise in plan--parallel geometry: in that case, once the population is updated at point $k$, the computation can be performed at point $k-1$ and will take exactly the new populations into account (cf. Fig. \ref{fig3}), since the radiation field at the innermost boundary is an input parameter of the model.

To make the algorithm computationally competitive, we proceed as follows. By analogy with the plan--parallel case, a way to implement the Gauss--Seidel algorithm would be not to account for the influence of the new populations $n_k^{new}$ on 
the incoming intensities $I_{k-1}^{in}$ and $I_{k}^{in}$ when we compute the outgoing intensities $I_{k}^{out}$. This would lead to performing 
\begin{itemize}
\item an update of $I_{k-1}^{out}$ using the new populations at points $k-2, k-1$, and $k$ using
$I_{k-2}^{out}$ and $I_{k-1}^{in}$ for the incoming intensities, calculated with the old populations.
\item an update of $I_{k}^{out}$ using the populations $n_{k-1}^{new}$, $n_{k}^{new}$, and $n_{k+1}^{old}$. As for the previous correction, the incoming intensities $I_{k}^{in}$ are given by the old populations.
\item an update of $I_{k+1}^{in}$ using the populations $n_{k}^{new}$, $n_{k+1}^{old}$, and $n_{k+2}^{old}$. 
\end{itemize}

As previously said, the first correction is in fact not necessary for speeding up the convergence rate. Actually, the mixing of \textit{old} and \textit{new} populations for the point $k$ reduces slightly the convergence rate (cf. section \ref{app_num}). Finally, we only perform the two latest corrections, which is equivalent to the corrections described by \citet{asensio2006}. 

\subsubsection{Upward and downward inversion}

As explained above, within the usual GS scheme and for each iteration, the grid is first swept
from $k=N$ to $k=1$ and then from $k=1$ to $k=N$, and during this second step, statistical equilibrium
equations are solved. As mentioned by \citet{trujillo1995} and \citet{trujillo1999},
a way to increase the convergence speed is to invert the statistical equilibrium
equations during both the upward and downward sweeping of the grid. 

During the downward propagation and when the iterative process has been initiated, we already know the intensities
$\{I_j^{out}\}_{j\in[1;N]}$. Thus, when reaching point $k$, we directly compute the averaged radiation field for 
$\theta \in [0;\pi]$. First we compute $I_k^{in}$ and then $I_k^{out}$ using
$I_k^{in}$ and $I_{k-1}^{out}$ as incoming intensities. Here we note that the intensities $I_{k-1}^{out}$ in fact involve a mixing 
of the populations obtained during the two last iterations. This is because the first correction 
described in the previous 
section is not performed and that the intensities are saved after the second correction. Finally, since the intensities $I_k^{out}$ are computed at this stage, there is only one correction
needed so that, after inversion, we perform an update of $I_{k}^{in}$ using the populations $n_{k+1}^{new}$, $n_{k}^{new}$, and $n_{k-1}^{old}$.
During the downward propagation, we have to save the intensities $I_{k}^{in}$ calculated with the new populations  since these intensities will be used during the upward propagation. 
The upward sweeping of the grid is similar to the one described in the previous section. 

\section{Numerical results} \label{app_num}

The numerical implementation was checked against the results obtained with the code described by
\citet{gonzalez1993}. Various models were used in the comparison, covering different physical conditions, as well as 
low, moderate, and large optical depths. The differences in the derived level populations ranged from 
a tenth of a percent to a few percent when integrated optical depths were higher than $\sim$ 100. This
corresponds to the expected accuracy of numerical radiative transfer codes as explained in \citet{zadelhoff2002}.

To accelerate the convergence we used the ALI method described in Sect. \ref{theorie:ALI} . 
In spherical geometry, the diagonal of the full
Lambda operator is given by Eqs. \ref{lamdiag1}, \ref{lamdiag2}, and \ref{lamdiag3}. 
However, determining the exact diagonal 
is time consuming: first, it requires the evaluation of the opacities $\Delta \tau_{m,m+l}^+$ that depend 
on all the internal shells crossed by the rays and second, it requires interpolating in angle the coefficients 
$c_{m+l}^{+l}$ to obtain them at the values $\theta_{m+l}^+$.  In practice, we thus chose not to keep the corresponding terms and we approximated 
the diagonal of the Lambda Operator just keeping, for $\theta > \pi/2$,
\begin{itemize}
\item if $m \geq K+1$ :
\begin{eqnarray}
 \Lambda_{m,m}  & = & 
 \beta_{m} (\theta_{m}^-) + \gamma_{m-1} (\theta_{m-1}^-) \, \textrm{e}^{-\Delta \tau_{m,m-1}^-}
\end{eqnarray}
\item if $ m = K $ :
\begin{eqnarray}
\Lambda_{m,m}  & = & 
  \alpha_m(\theta_m^-)+\beta_{m} (\theta_{m}^-)
+ \left[ \beta_{m}(\theta_m^+) + \gamma_{m} (\theta_{m}^+) \right] \textrm{e}^{-\Delta \tau_{m,m}^+}
\qquad .
\end{eqnarray}
\end{itemize}
For $\theta \leq \pi/2$, we conserve the terms given by expressions \ref{lamb1a} and \ref{lamb1b}.

\begin{figure}[h!]
   \centering
   \includegraphics[scale=0.35,angle=270]{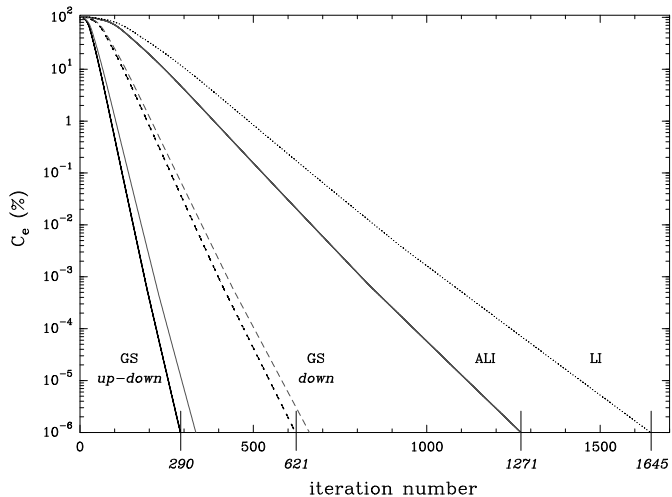}
      \caption{Convergence properties of the various algorithms in terms of number of iterations.}
         \label{fig4}
\end{figure}

These terms permit us to accelerate the rate of convergence notably by comparison to the LI scheme. Moreover,
the approximation made in deriving the diagonal of the ALO does not induce false convergence. Both aspects can be seen in Fig. \ref{fig4} where the maximum difference with respect to the converged solution is plotted. The model considered here corresponds to a sphere of uniform physical parameters with n(H$_2$) = $10^4$ cm$^{-3}$ and temperature T $= 25 K$. The sampling consists in $\sim$ 500 spatial grid points and 20 angles. This has been found to insure a convergence for the level populations better than 0.1 \% for the first 12 levels. The molecule considered is ortho--H$_2$CO, and 14 energy levels were included in the calculation. 
The molecular parameters were taken from the references given in Table \ref{tab1}.
Since no analytical solution exists for this problem, we used as reference populations those obtained with the LI scheme.  The error relative to the converged solution for the considered grid, noted $C_e$, is defined according to Eq. 20 of \citet{auer1994}.

\begin{table*}
\centering
\begin{tabular}{|c||c|c|c|}
\hline
molecule   & rate coefficients & dipole moment & energy levels \\
\hline \hline
H$_2$CO    & \citet{green1991}   & \citet{fabricant1977} & \citet{muller2005} \\
HCN        & \citet{monteiro1986}& \citet{ebenstein1984} & \citet{muller2005} \\
HNC        & \citet{monteiro1986}& \citet{blackman1976}  & \citet{muller2005} \\
N$_2$H$^+$ & \citet{daniel2005}  & \citet{havenith1990}  & \citet{caselli1995} \\
\hline
\end{tabular}
\caption{References for the molecular parameters used in this work.}
\label{tab1}
\end{table*}

Figure \ref{fig4} also shows the rate of convergence of the different algorithms in terms of the iteration number. For the GS algorithms, two curves are shown: the grey one corresponds to the case where the incoming intensities are corrected (i.e. correction 1 of Sect. \ref{upward_inv}) and the black one corresponds to the case where this correction is omitted.
 We see that, for this model, the ALI scheme speeds up the convergence by a factor 1.3 in comparison to the LI scheme.  
The GS down and GS up-down enabled us to reduce the number of iterations by factors of 2.0 and 4.4 respectively by comparison to the ALI scheme. In Sect. \ref{theorie:GS}, the various corrections to be done for these two algorithms were discussed. 
For each correction, the code calls a formal solver that takes the source functions of the lines as input and outputs the averaged radiation field and Lambda operator, as well as the intensities. Thus, considering the number of calls to the formal solver and in the different schemes, we have for each layer 2 calls in the ALI scheme, 4 calls in the GS down scheme, and 6 calls in the GS up--down scheme. The number of calls gives the relative time per iteration of each algorithm. In practice, the GS schemes implementation can be improved by performing some bookkeeping, in particular for the incoming intensities. 
These lead to times per iteration, in the GS down and GS up-down schemes, which are respectively 1.45 and 2.3 times 
greater than in the ALI scheme. These times were corrected for the time spent in the code in internal storage on  
the hard drive. Such a procedure is done in our implementation of the GS schemes while not for the ALI one and makes the GS scheme $\sim$ 20\% slower than permitted by the algorithmic. Moreover, these longer times are compensated for by the reduced number of iterations required (see above). Thus, for the current implementation, the GS down and GS up-down respectively reduce the total calculation time by factors of 1.4 and 1.9. The numerical implementation of the GS scheme could still be improved to gain some computing time.  The corrections in spherical geometry are formally similar to the one in plan--parallel geometry, and for this case, \citet{paletou2007} find that the time per iteration of the GS down scheme was only 1.3 longer than that of the ALI scheme. Our code has been written in a very modular and structured way. By decreasing the number of internal calls to routines and functions, we could still gain computing time and approach the 1.3 value of plan-parallel geometry.  

\begin{figure}[h!]
   \centering
   \includegraphics[scale=0.35,angle=270]{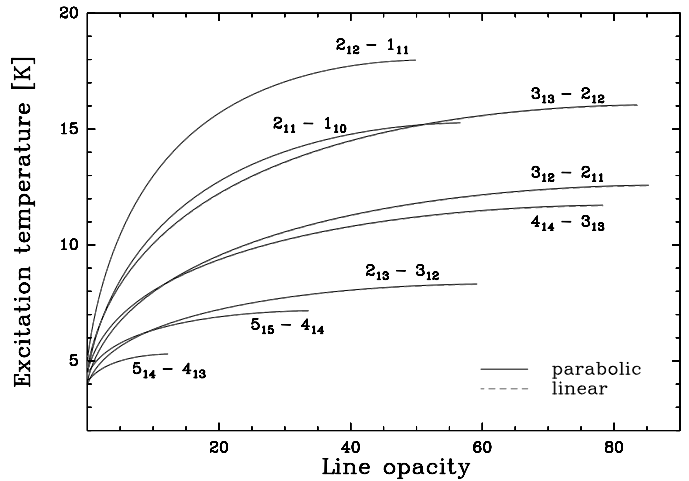}
      \caption{Excitation temperature versus opacity for a model considering ortho--H$_2$CO.}
         \label{fig5}
\end{figure}

Nevertheless, the GS schemes become more advantageous if, rather than parabolic interpolation, we use linear interpolation.  In that case, there are fewer corrections to carry out so that the time per iteration of the GS down and GS up-down scheme are only 1.4 and 1.7 that of the ALI scheme.  The drawback is that if using linear interpolation, one needs to use more grid points in order to converge to the true solution. Nevertheless, for the model considered here, we find that the differences in level populations are inferior to 1\%. In Fig. \ref{fig5}, the line parameters of the most opaque lines are plotted for the two types of interpolations and the results are indistinguishable on the scale of the graph.

We note that the various algorithms discussed here can still be improved if combined with the Ng 
acceleration technique \citep{ng1974}. For the various models tested, we found that it typically leads to a 
reduction by a factor of 3 in the number of iterations. Alternatively, the successive over--relaxation technique 
can be used \citet{trujillo1995}.

\section{Overlap of hyperfine lines}\label{app_LO}

\begin{figure}
\centering
   \includegraphics[scale=0.35,angle=270]{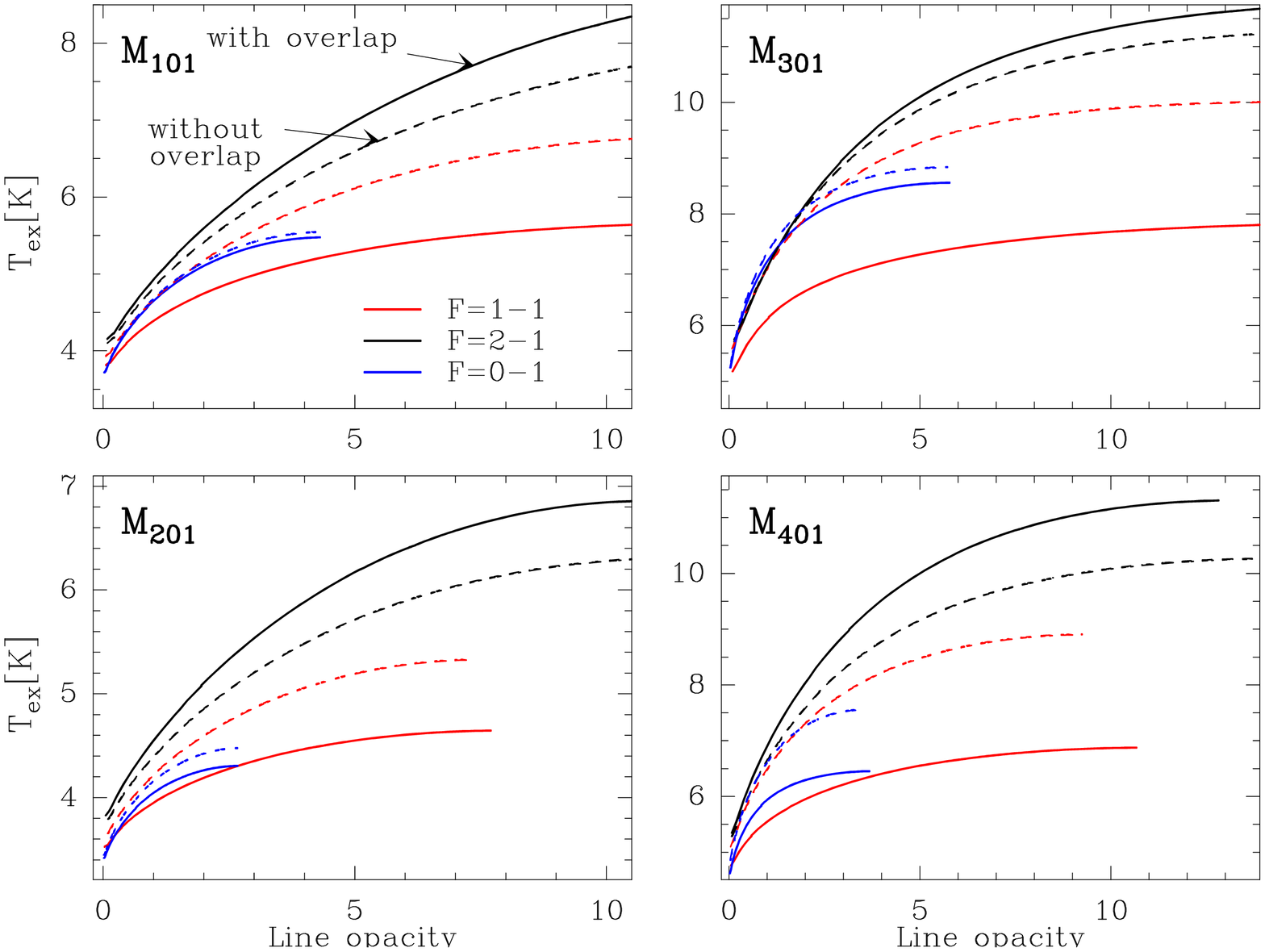}
   \caption{Line parameters obtained for the models described in \citet{gonzalez1993}}
   \label{fig6}
\end{figure}

The numerical implementation of the overlap was checked according to the results 
reported by \citet{gonzalez1993}. For the molecules considered in what follows the molecular parameters are summarised in Table \ref{tab1}.
Figure \ref{fig6} shows the variation in excitation temperature versus line opacity 
for the three HCN hyperfine lines associated with the $J=1-0$ rotational transition.  
The physical parameters of the models were taken from \citet{gonzalez1993} 
. The line parameters are given with (solid lines) 
and without (dashed lines) introduction of the overlap.
By comparison to the results reported in 
Figs. 1a--1d of \citet{gonzalez1993}, we see that the qualitative behaviour of $T_{ex}$ is correctly
reproduced for all the hyperfine lines when the overlap is introduced. Nevertheless, 
a quantitative comparison of the line parameters, with and without overlap, 
shows differences in the order 
of 5--10\%. These differences are due 
to the choice of grid parameters, especially the number of radial grid points, adopted to reach convergence. 

\begin{figure}
\centering
   \includegraphics[scale=0.6,angle=0]{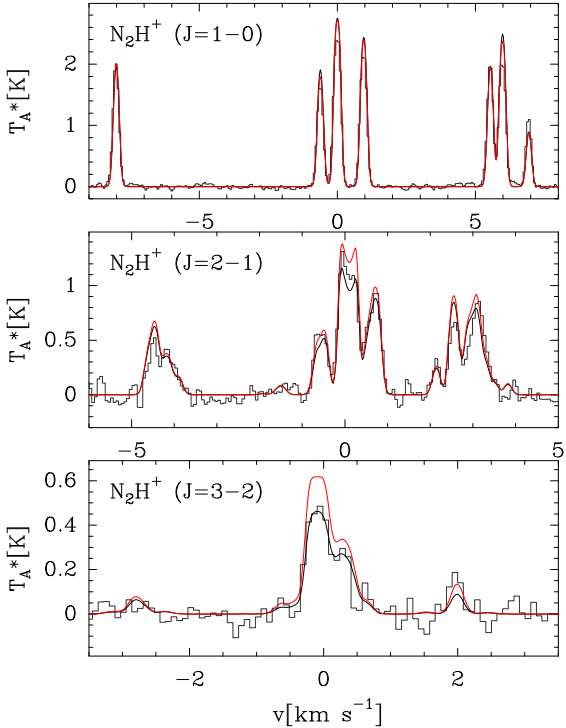}
   \caption{Effect of the overlap on the line profile of L63. The black line correspond 
     to a model without overlap and the red lines correspond to models that include overlap.}
   \label{fig7}
\end{figure}

Figure \ref{fig7} shows the influence of the overlap on the modelization 
of the hyperfine lines associated with the 3 lowest N$_2$H$^+$ rotational transitions.
This molecule is particularly interesting for infering the physical conditions of the densest 
parts of cold dark clouds. Indeed, it was found that this molecule does not suffer from strong 
depletion for H$_2$ volume densities below $\sim 10^6$ cm$^{-3}$. 
The model
parameters are identical to those reported in \citet{daniel2007} for the source L63.
In this figure, we see that the $J=1-0$ line remains mainly unchanged with differences in the line fluxes
lower than 5\%. The differences increase when considering higher rotational lines.
For the $J=3-2$ line, including the overlap increases the line flux by $\sim$ 20-30 \%.    
This can be understood as follows. For a group of blended hyperfine lines, the average radiation field seen by 
each line is greater when considering the overlap, since more molecules can interact
with the radiation. Consequently, the pumping of the upper level of each transition 
is more efficient, which results in an enhancement of the populations 
of the highest rotational levels. Therefore the overlap changes 
the physical parameters reported in \citet{daniel2007}. Mainly,
H$_2$ volume densities and/or temperatures were overestimated in this study.

\begin{figure*}
\centering
   \includegraphics[scale=0.9,angle=0]{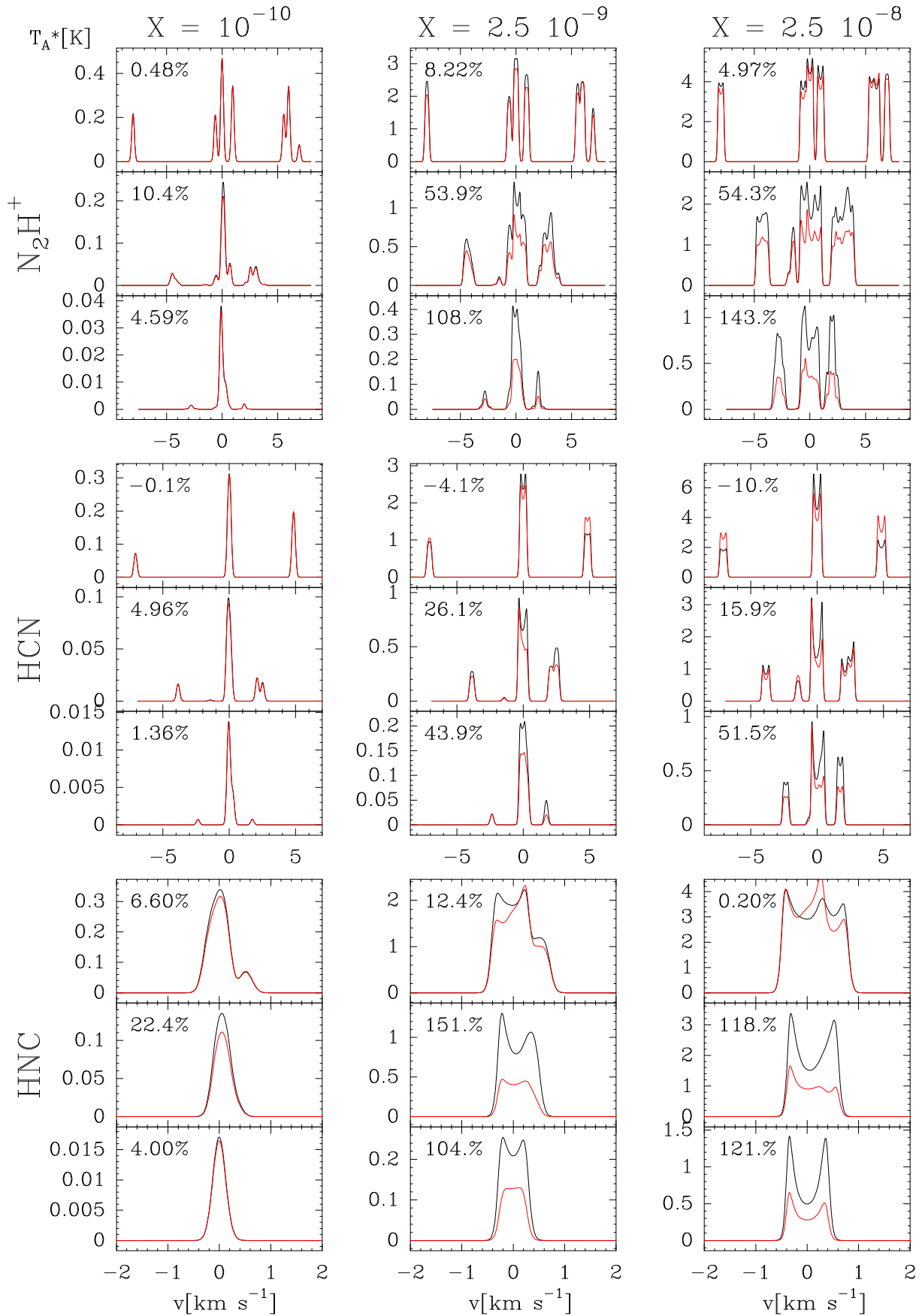}
   \caption{Effect of overlap on the N$_2$H$^+$, HCN, and HNC $J=1-0$, $2-1$, and $3-2$ lines.
    Red lines correspond to models without introduction of the overlap and black lines to models considering line overlap. The differences between integrated intensities are given for each line. The intensity scale is antenna temperature and are obtained using the standard values of HPBW and beam efficiency of the 30m IRAM telescope.}
   \label{fig8}
\end{figure*}

\begin{figure}
\centering
   \includegraphics[scale=0.6,angle=0]{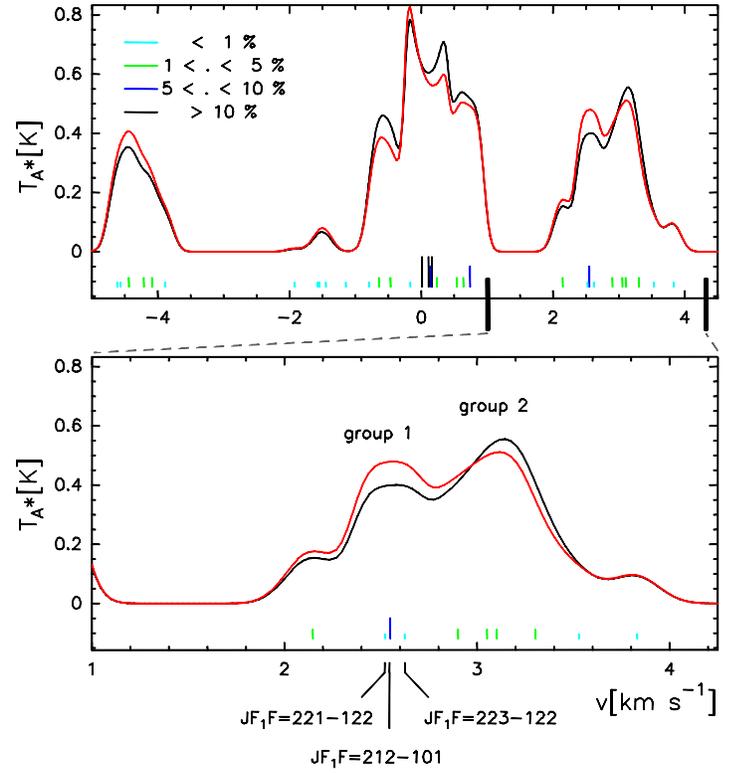}
   \caption{Effect of overlap on relative line intensities. The black curve corresponds
to a model that includes line overlap and the red line to a model without overlap, the latter
being scaled by the integrated intensity ratio.}
   \label{fig9}
\end{figure}

The effect induced by line overlap obviously depends on individual hyperfine line opacities, 
as well as on the frequency offset between the lines. Figure \ref{fig8} shows the emerging 
intensities for the three lowest rotational lines of HCN, HNC, and N$_2$H$^+$. The models consist
of a uniform density core of diameter 1'. The density is n(H$_2$) = $10^5$ cm$^{-3}$, 
the temperature T=10 K, and the turbulence is set to v$_{turb}$ = 0.15 km.s$^{-1}$. 
For each molecule, calculations are performed at abundances of 
$10^{-10}$, $2.5 \cdot 10^{-9}$, and $2.5 \cdot 10^{-8}$. First, it can be seen that, at the lowest abundance, 
the effect introduced by the overlap is weak for N$_2$H$^+$ and HCN. The effect is stronger 
for HNC due to closer hyperfine frequencies.
Second, we find that the effect introduced by the overlap is approximatively proportional to the column density, as long as individual line opacities $\leq 1$. In the present calculations, this is found for the $J=3-2$ lines of each species for the 2 lowest abundances. For these lines, we see that the integrated intensity ratios of the calculations with and without overlap are approximatively proportional to the abundance ratio. In a more optically thick regime, the increase in the integrated intensity ratio depends less on the column density and finally reaches a saturation regime where the intensity ratio starts to be affected by self--absorption effects. This can be seen in Fig. \ref{fig8} for the $J=2-1$ and $J=1-0$ lines. Finally, the error introduced in the interpretation of observations when neglecting the line overlap can be especially important  for HNC. Indeed, we see in Fig. \ref{fig8} that, for conditions typical of dark clouds, including line overlap leads to an intensity increase of up to a factor 2. Moreover, for similar abundances, the overlap differentially affects HCN and HNC and thus, neglecting it may lead to overestimating the X(HNC)/X(HCN) ratio.

\begin{figure}
\centering
   \includegraphics[scale=0.35,angle=270]{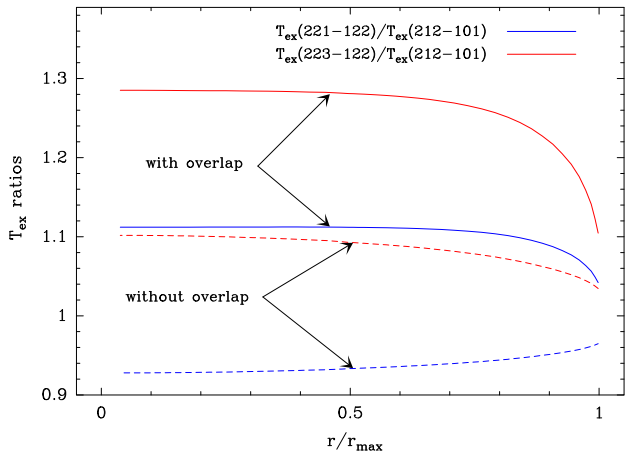}
   \caption{Excitation temperatures ratios for overlapping lines with different line strengths.}
   \label{fig10}
\end{figure}

Another effect introduced by the overlap is to modify the relative intensities of the hyperfine components of a given rotational transition. Figure \ref{fig9} shows the $J=2-1$ line of N$_2$H$^+$ for a model with abundance 2.5 10$^{-9}$. The line obtained without overlap is represented multiplied by a factor of 1.54, which corresponds to the integrated intensity ratio obtained previously. In this figure the position in frequency of the different hyperfine components is also represented as are their line strengths. These are given in percent by reference to the value obtained by summation over all the hyperfine components. Considering the red satellite, we see that, when including the overlap, the intensity associated with the strongest transition (group 1 at v $\sim$ 2.5 km s$^{-1}$) is reduced by comparison to the group of weaker lines situated at v $\sim$ 3 km s$^{-1}$ (group 2). 
This effect on the profile for the N$_2$H$^+$ $J=2-1$ red satellite agrees with the observations reported in \citet{daniel2007}.  
Two effects yield to the current emerging profile.
The main effect is introduced by the number of strong transitions that overlap in each group.
In group 1, the line strengths of the two weakest transitions are low enough so that there is mostly one line that contributes to the emission. On the other hand, group 2 corresponds to 4 lines with equivalent line strengths. Thus, for this group of lines, the pumping of the upper levels is enhanced producing an increase in the emergent intensities.
Another effect of the overlap is to create a flow of population that depends on the relative line strengths of the overlapping lines. Figure \ref{fig10} shows excitation temperature ratios
between the transitions of group 1 obtained with (solid lines) and without (dashed lines) introduction of the overlap. We see in this figure that the overlap enhances the excitation temperatures of the 2 weakest lines. This means that the levels $JF_1F$ = 223 and $JF_1F$ = 221 are more efficiently populated when
considering the overlap due to the presence of the $JF_1F=212-101$ transition that is close in frequency.
This can be understood when considering the way the average radiation field is obtained when solving the radiative transfer. For the strongest line, the presence of weak hyperfine components close in frequency does not have a strong influence since 
the opacities/intensities in the wings of the intrinsic line profiles are low. The average radiation field is thus only slightly modified. On the other hand, for a weak line that is close to the frequency of a strong line, the average radiation field is increased. This will pump the corresponding energy levels and result in an increase in intensity for the weakest lines.

\section{Conclusions}

We have presented technical details concerning the determination
of the Lambda operator. Moreover, we gave details on 
the implementation of the Gauss--Seidel
algorithm in spherical geometry. Finally, the numerical code was applied
to the treatment of line overlap.
The main conclusions are:
   \begin{enumerate}
      \item The diagonal of the Lambda operator, as obtained within
            the short characteristics method, has to be approximated in practical applications.
            The use of approximated diagonal terms still enable speeding the convergence. 
            Moreover, the tests carried out did not show false convergence by comparison with LI calculations. 
      \item The implementation of the Gauss--Seidel algorithm in spherical geometry also requires
            some approximations. These are introduced by analogy with the plan--parallel case.
      \item The convergence rate of the GS down and GS up--down are similar to what is obtained in plan--parallel geometry, i.e. they reduce the number of iterations to reach convergence by factors of 2 and 4 respectively. The gain in terms of computing time strongly depends on the numerical implementation, and in the present code, the GS up--down reduces the calculation time by a factor $\sim$ 2 by comparison to the ALI scheme.
       \item Line overlaps can lead to a large increase in the predicted intensities, the effect being greater for higher rotational lines. Especially, differential effects for HCN and HNC can alter the estimate of the abundance ratio X(HNC)/X(HCN), and models where line overlap is neglected will overestimate it. 

    \end{enumerate}

\begin{acknowledgements}
      Part of this work was supported by the AYA2006-14876, ESP2004-00665, ESP2007-65812-C02-01, ESO nº 010103050002, Molecular Universe MRTN-CT-2004-51230, ASTROCAM S-0505 ESP-0237 projects. The authors thank F. Paletou and L. L\'eger for the
help provided concerning the Gauss--Seidel algorithm. We also thank J.R Pardo for his useful revision of the 
present manuscript. 

\end{acknowledgements}

\begin{appendix} 

\section{Discretized Lambda operator: plan--parallel geometry}

As pointed out by \citet{rybicki1991}
the Lambda operator is in fact an approximation of the true Lambda operator since it depends
on the discretization of the structure of the object. Thus, the Lambda operator is a square matrix 
whose dimension corresponds to the number of spatial grid points.  

Within the SC scheme, the Lambda operator can be constructed easily. As a first step,
from Eqs. \ref{eq1} and \ref{eq2} one finds that along a characteristic
\begin{eqnarray}
\begin{array}{cc}
I_i = I_0 \, \textrm{e}^{-\tau_{i}} + & \underbrace{ 
\sum_{k=0}^{i-1} \Delta I_{i-k} \, \textrm{e}^{-\Delta \tau_{i,i-k}}} \\
 & \tilde{\Lambda}^i \\
\end{array}
\end{eqnarray}
where $\Delta \tau_{i,j} = \tau_{i} - \tau_{j}$. To obtain a compact expression, we adopt from now on the notation $a_i^{-1} = \alpha_i$, $a_i^0 = \beta_i$,
and $a_i^1 = \gamma_i$ for the terms that appear in Eq. \ref{eq2}. Introducing it in the previous expression,
we find
\begin{eqnarray}
\displaystyle \tilde{\Lambda}^i = 
\sum_{k=0}^{i+1} \left(  \sum_{l=-1}^{1} \tilde{b}_{k+l}^{-l} \, 
\textrm{e}^{-\Delta \tau_{i,k+l}}  \right) S_\nu (\tau_k) 
\label{geom:lambda_tot}
\end{eqnarray}

\begin{eqnarray}
\textrm{with} \quad \tilde{b}_{k+l}^{-l} = \left\{
\begin{array}{l}
\displaystyle
 a_{k+l}^{-l} \quad \textrm{if} \quad  0 < k+l \leq i \vspace{0.2cm} \\
0 \quad \textrm{otherwise} \qquad . \\ 
\end{array}
\right. 
\label{geom:condition1}
\end{eqnarray} 
In the above expression, the source functions are given at optical depths
that correspond to the intersection of the ray with the discretized 
structure of the cloud. In order to construct the Lambda operator, it is then
necessary to introduce a notation where the source functions are 
given as a function of grid points.

\begin{figure}[h!]
   \centering
   \includegraphics[scale=0.3]{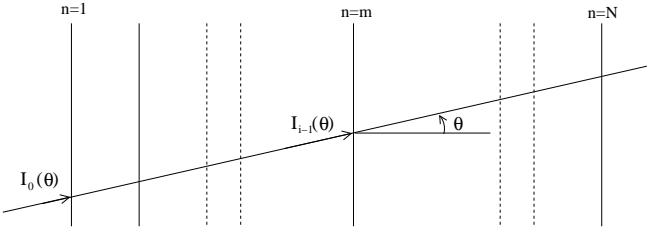}
      \caption{Notation adopted in plan--parallel geometry.}
         \label{fig11}
\end{figure}

In plan--parallel geometry, the values for $S_\nu (\tau_k)$ are readily assigned to values of 
the source functions for each layer. 
From Fig. \ref{fig11}, we have in the case $\theta \in [0;\pi/2]$ that the 
point $ k = 0$ in expression \ref{geom:lambda_tot} corresponds to the layer $n = 1$. The point $k = i-1$ 
corresponds to the layer $n = m$. In the above expression, we thus first introduce the variable 
$n = k+1$ and then a notation where all the quantities are indexed according
to the layer, i.e.
\begin{eqnarray}
\begin{array}{lcl}
S_\nu(\tau_{n-1}) & \to & S_\nu^n  \\
\tilde{b}_{n-1+l}^{-l}               & \to & b_{n+l}^{-l} \\
\Delta \tau_{n-1,n-1+l} & \to &  \Delta \tau^-_{n,n+l} \quad . \\  
\end{array}
\end{eqnarray}
The condition \ref{geom:condition1} transforms as
\begin{eqnarray}
b_{n+l}^{-l} = \left\{
\begin{array}{l}
\displaystyle
 a_{n+l}^{-l} \quad \textrm{if} \quad  1 < n+l \leq m+1 \vspace{0.2cm} \\
0 \quad \textrm{otherwise} \quad ,\\ 
\end{array}
\right. 
\end{eqnarray} 
where the coefficient $a_{n+l}^{-l}$ keeps the same definition given above, but is
now labelled according to the layer index.
It leads to 
\begin{eqnarray}
\begin{array}{rclc}
\displaystyle \tilde{\Lambda}^{\; i-1} =  \Lambda^{m} 
= \sum_{n=1}^{m+1} & \underbrace{\left(  \sum_{l=-1}^{1} b_{n+l}^{-l} \, 
\textrm{e}^{-\Delta \tau^-_{m,n+l}}  \right)}& S_\nu^n  \quad ,\\
& \Lambda_{m,n} & & 
\end{array}
\end{eqnarray}
In this expression,  $\Lambda_{m,n}$ stands for the element of the $m^{th}$ line
and $n^{th}$ column of the Lambda operator when $\theta \in [0;\pi/2]$.

Equivalent transformations can be done in the case $\theta \in [\pi/2;\pi]$. Defining N so that $n = N-k$,
identifying $i = N-m$
and introducing the transformation
\begin{eqnarray}
\begin{array}{lcl}
S_\nu(\tau_{N-n}) & \to & S_\nu^n  \\
\tilde{b}_{N-(n-l)}^{-l}               & \to & b_{n-l}^{-l} \\
\Delta \tau_{N-n,N-(n-l)} & \to &  \Delta \tau^-_{n,n-l} \quad ,\\  
\end{array}
\end{eqnarray}
with
\begin{eqnarray}
 b_{n+l}^{-l} = \left\{
\begin{array}{l}
\displaystyle
 a_{n+l}^{+l} \quad \textrm{if} \quad  m \leq n+l < N \vspace{0.2cm} \\
0 \quad \textrm{otherwise} \quad ,\\ 
\end{array}
\right. 
\end{eqnarray} 
leads to 
\begin{eqnarray}
\displaystyle \Lambda^m = \sum_{n=m-1}^{N} \left(  \sum_{l=-1}^{1} b_{n+l}^{+l} \, 
\textrm{e}^{-\Delta \tau^-_{m,n+l}}  \right)  S_\nu^n  \quad .
\end{eqnarray}
This final expression defines the element $\Lambda_{m,n}$ when $\theta \in [\pi/2;\pi]$.

\section{Discretized Lambda Operator: spherical geometry}

In spherical geometry, the same procedure of the plan--parallel case 
can be followed to obtain the components of the Lambda operator.
In this case, however, 
all the shells are not necessarily crossed for a given direction, and furthermore, some shells are crossed 
two times before reaching the current point. 

\begin{figure}[h!]
   \centering
   \includegraphics[scale=0.4]{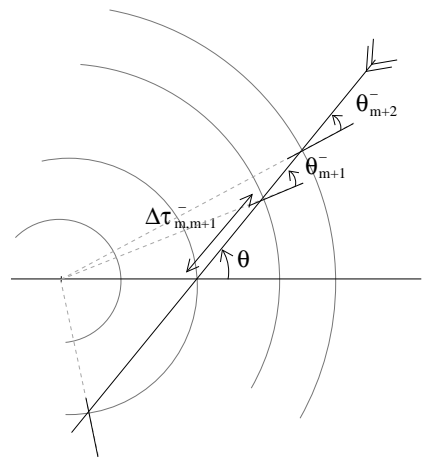}
      \caption{Definition of geometrical quantities.}
         \label{fig12}
\end{figure}

We consider a cloud discretized in $N$ concentric shells. In the case 
$\theta \in [0;\pi/2]$, the transformations are similar to the second case 
examined in the previous section with two exceptions. First, in the SC frame,
the coefficients $b_{n+l}^{+l}$ depend on an angle 
defined in a coordinate system associated to the layer $n+l$. This dependence was not 
outlined in the plan--parallel case since all the coefficients along a characteristic are
at a unique angle value, namely $\theta$. In what follows, we introduce the notation  
$\theta^{\pm}_{n+l}$, which refers to the angle to be taken into account in the evaluation of $b^{+l}_{n+l}$ (see Fig. \ref{fig12}).
The superscript in this notation distinguishes acute (-) and obtuse (+) angles.  

Secondly, depending on the value of $\theta$, 
the source function corresponding to the point $i+1$ can be associated either to shell $m$ or 
to shell $m-1$. In this case, we obtain for
$\theta \in [0;\pi-\xi_{m,m-1}]$ 
\begin{eqnarray}
\displaystyle \Lambda^m = \sum_{n=m-1}^{N} \left(  \sum_{l=-1}^{1} 
b_{n+l}^{+l} (\theta_{n+l}^-)  \, 
\textrm{e}^{-\Delta \tau^-_{m,n+l}}  \right)  S_\nu^n \quad ,
\label{lamb1a}
\end{eqnarray}
and, for $\theta \in [\pi-\xi_{m,m-1};\pi/2]$ 
\begin{eqnarray}
\displaystyle \Lambda^m = \sum_{n=m}^{N} \left(  \sum_{l=-1}^{1} 
b_{n+l}^{+l} (\theta_{n+l}^-)  \, 
\textrm{e}^{-\Delta \tau^-_{m,n+l}}  \right)  S_\nu^n
+ \gamma_{m} (\theta_m^-) \, S_\nu^m \quad .
\label{lamb1b}
\end{eqnarray}

\begin{figure}[h!]
   \centering
   \includegraphics[scale=0.4]{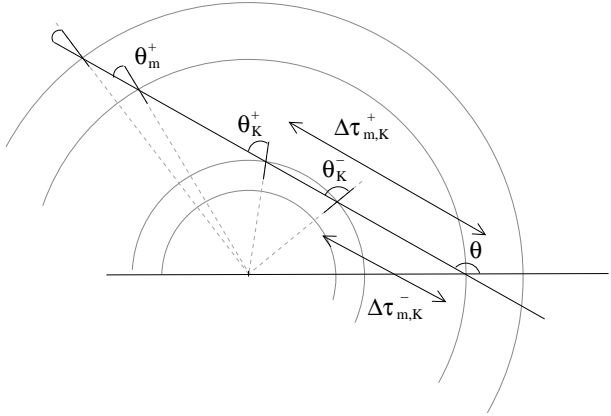}
      \caption{Definition of geometrical quantities.}
         \label{fig13}
\end{figure}

In the above expressions, we have introduced the notation $\Delta \tau^{\pm}_{m,n}$. 
A ray that intersects the shells $m$ and $n$ defines two segments. The opacity along the longest
one defines $\Delta \tau^{+}_{m,n}$, and the opacity along the smallest one defines 
$\Delta \tau^{-}_{m,n}$ (see Figs. \ref{fig12} and \ref{fig13}).

Now, let us consider the case $\theta > \pi/2$, and designate $K$ the innermost shell that is intersected 
by the ray. This shell is such that we have    $\xi_{m,K} < \theta < \xi_{m,K-1}$,
with the angle $\xi_{m,K}$ defined according to Eq. \ref{angle}.

Manipulations of Eq. \ref{geom:lambda_tot} lead to
\begin{eqnarray}
\Lambda^m = \sum_{k=K}^N \Lambda_{m,k} \, S^k 
\end{eqnarray}
with
\begin{itemize}
\item if $K+1 < k \leq m+1$ :
\begin{eqnarray}
 \Lambda_{m,k}  =  \mathcal{A}_{m,k}\left( \theta \right)
\end{eqnarray}
\item if $ m+1 < k \leq N$ :
\begin{eqnarray}
 \Lambda_{m,k}  =  \sum_{l=-1}^1 c_{k+l}^{+l} \, \textrm{e}^{-\Delta \tau_{m,k+l}^+ }
\end{eqnarray}
\item if $ k = K+1 $ :
\begin{eqnarray}
 \Lambda_{m,K+1}  =  \mathcal{A}_{m,K+1}\left( \theta \right)
+ \alpha_K(\theta_K^+)  \textrm{e}^{-\Delta \tau_{m,K}^+ } \nonumber + 
\gamma_K(\theta_K^-) \, \textrm{e}^{-\Delta \tau_{m,K}^- }
\end{eqnarray}
\item if $ k = K$ :
\begin{eqnarray}
 \Lambda_{m,K}  & = &  \mathcal{A}_{m,K}\left( \theta \right)
+ \beta_K(\theta_K^+)  \textrm{e}^{-\Delta \tau_{m,K}^+ } + 
\left[ 
\left( \alpha_K(\theta_K^-) + \beta_K(\theta_K^-) \right) \textrm{e}^{-\Delta \tau_{m,K}^- } \right. 
 \nonumber \\ & & \left. 
+ \gamma_K(\theta_K^+)  \textrm{e}^{-\Delta \tau_{m,K}^+ }
\right] \times
\left( 
1 - \delta( \theta - \xi_{m,K}) 
\right) \quad .
\end{eqnarray}
\end{itemize}
The quantity $\mathcal{A}_{m,k}\left( \theta \right)$ is defined according to 
\begin{eqnarray}
\mathcal{A}_{m,k}\left( \theta \right) =
\sum_{l=-1}^1 \left[ c_{k+l}^{+l} (\theta_{k+l}^+) \, \textrm{e}^{-\Delta \tau_{m,k+l}^+ }
+ b_{k+l}^{-l} (\theta_{k+l}^-) \, \textrm{e}^{-\Delta \tau_{m,k+l}^- } \right]
\end{eqnarray}
\begin{eqnarray}
\textrm{with} \quad c_{k+l}^{+l} = \left\{
\begin{array}{l}
\displaystyle
 a_{k+l}^{+l} \quad \textrm{if} \quad  K < k+l < N  \vspace{0.2cm} \\
0 \quad \textrm{otherwise} \quad , \\ 
\end{array}
\right. \\
\textrm{and} \quad b_{k+l}^{-l} = \left\{
\begin{array}{l}
\displaystyle
 a_{k+l}^{-l} \quad \textrm{if} \quad  K < k+l \leq m  \vspace{0.2cm} \\
0 \quad \textrm{otherwise} \quad .\\ 
\end{array}
\right.
\end{eqnarray} 

In the expressions derived above, the dependence on the variable $\theta$ is not made explicit 
but would appear in the definitions of $\theta_{k+l}^{\pm}$ and 
$\Delta \tau_{m,k+l}^{\pm}$.
The expressions obtained for $\Lambda_{m,k}$ are suitable for constructing an
operator of whatever bandwidth. However, we are interested here
in the diagonal of this operator: 

\begin{itemize}
\item if $m > K+1$ :
\begin{eqnarray}
 \Lambda_{m,m}  & = & 
 \sum_{l=-1}^1  c_{m+l}^{+l} (\theta_{m+l}^+) \, \textrm{e}^{-\Delta \tau_{m,m+l}^+ }
 +  \beta_{m} (\theta_{m}^-) \nonumber \\ & & 
+ \gamma_{m-1} (\theta_{m-1}^-) \, \textrm{e}^{-\Delta \tau_{m,m-1}^-}
\label{lamdiag1}
\end{eqnarray}
\item if $ m = K+1 $ :
\begin{eqnarray}
 \Lambda_{m,m} & = & 
 \sum_{l=0}^1  c_{m+l}^{+l} (\theta_{m+l}^+) \, \textrm{e}^{-\Delta \tau_{m,m+l}^+ }
 +  \beta_{m} (\theta_{m}^-) \nonumber \\ & & 
+ \gamma_{m-1} (\theta_{m-1}^-) \, \textrm{e}^{-\Delta \tau_{m,m-1}^-}
 + \alpha_{m-1} (\theta_{m-1}^+) \, \textrm{e}^{-\Delta \tau_{m,m-1}^+}
\label{lamdiag2}
\end{eqnarray}
\item if $ m = K $ :
\begin{eqnarray}
\Lambda_{m,m}  & = & 
 \sum_{l= 1}^1  c_{m+l}^{+l} (\theta_{m+l}^+) \, \textrm{e}^{-\Delta \tau_{m,m+l}^+ } 
 +  \alpha_m(\theta_m^-)+\beta_{m} (\theta_{m}^-) \nonumber \\ & & 
+ \left( \beta_{m}(\theta_m^+) + \gamma_{m} (\theta_{m}^+) \right) \textrm{e}^{-\Delta \tau_{m,m}^+}
\quad .
\label{lamdiag3}
\end{eqnarray}
\end{itemize}

\end{appendix}

\end{document}